\documentclass{elsart1p}

\usepackage{graphics}
\usepackage{graphicx}
\usepackage{dcolumn}
\usepackage{bm}
\usepackage{float}
\usepackage{epsfig}
\usepackage{stmaryrd}
\usepackage{latexsym}
\usepackage{amssymb}
\usepackage{amsfonts}
\usepackage{amsmath}

\journal{Solid State Communications}

\newcommand{\prl}{Phys. Rev. Lett. }

\newcommand{\pra}{Phys. Rev. A }
\newcommand{\prb}{Phys. Rev. B }

\begin{document}
\begin{frontmatter}

\renewcommand{\thefigure}{\arabic{figure}}

\title{Analytic theory of pair distribution functions in symmetric electron-electron and electron-hole bilayers}

\author[SNS]{S. H. Abedinpour},
\author[IPM]{R. Asgari},
\author[SNS]{M. Polini\corauthref{cor}},
\corauth[cor]{Tel.: +39 050 509038; Fax: +39 050 563513.}
\ead{m.polini@sns.it}
\author[SNS]{M. P. Tosi}
\address[SNS]{NEST-CNR-INFM and Scuola Normale Superiore, I-56126 Pisa, Italy}
\address[IPM]{Institute for Studies in Theoretical Physics and Mathematics, Tehran 19395-5531, Iran}
\date{\today}

%

\begin{abstract}
We present a self-consistent analytic theory of the intra-layer and inter-layer pair correlation functions in electron-electron and electron-hole fluid bilayer systems. Our approach involves the solution of a zero-energy scattering Schr\"{o}dinger equation with an effective potential which includes a Bose-like term from Jastrow-Feenberg correlations and a Fermi term from kinetic energy and exchange, tailored to yield the Hartree-Fock limit at high carrier density. The theory is also shown to satisfy the plasmon sum rule and the charge neutrality condition. We obtain good agreement with the available Quantum Diffusion Monte Carlo (DMC) data in symmetric bilayers over a wide range of carrier density and layer spacing, and stress the role of three-body correlation terms in yielding fully quantitative agreement at strong coupling. Signals of impending transitions to density-modulated states at strong coupling and low layer spacing appear in the calculated pair correlations through inter-layer in-phase oscillations for electron-hole bilayers and out-of-phase oscillations for electron-electron bilayers, in agreement with the DMC findings.
\end{abstract}


\end{frontmatter}
\maketitle

\section{INTRODUCTION}
Advances in nanoscale growth technology have for a number of years allowed the fabrication of materials formed by alternating layers of two or more semiconductors. Electron or hole layers trapped at the various interfaces have made available Fermi-liquid systems having a quasi-two-dimensional (2D) low-energy dynamics under good control of carrier density and inter-layer spacing~\cite{ref:yamada}. It was soon realized that systems of spacially separated carriers present some advantages over conventional samples in which the carriers occupy the same region. Thus in electron-electron bilayer systems (EEB) evidence has been reported for novel fractional quantum Hall states~\cite{ref:fQHE} and for a bilayer quantum Wigner solid~\cite{ref:neilson_91,ref:shayegan_96}. In electron-hole bilayer systems (EHB), on the other hand, the Coulomb attractions between the two kinds of carriers can bring about inter-layer coherence and hence again crystalline order as well as exciton condensation and superfluidity~\cite{ref:keldish}.

Many of the Coulomb interaction effects in these systems can be understood with the help of the homogeneous 2D electron gas model~\cite{ref:vignale_book}. A central role in the theory is played by the intra-layer and inter-layer pair correlations as expressed by the instantaneous pair distribution functions $g_{\alpha\beta}(r)$, where a Greek suffix $\alpha=1$ or $2$ stands for carriers in one of the two layers. These functions are defined so that the quantity $2\pi r n_{\beta}g_{\alpha\beta}(r)dr$ gives the average number of particles lying at mean density $n_{\beta}$ in a shell of radius $r$ and thickness $dr$ inside layer $\beta$, centered on the in-plane position of a particle in layer $\alpha$. Quantum simulation studies of these functions by DMC methods have been reported for both EEB~\cite{ref:rapisarda,ref:moudgil}  and EHB~\cite{ref:depalo_02,ref:moudgil} symmetric systems at densities $n_1=n_2=n$ and masses $m_1=m_2=m$, for a variety of values of the inter-layer distance $d$ and of the coupling strength $r_s$ [here defined as $r_s=(a_{\rm B}\sqrt{n\pi})^{-1}$ where $a_{\rm B}$ is the Bohr radius]. Theoretical calculations of $g_{\alpha\beta}(r)$ based on a quantum STLS scheme and comparisons with DMC data have been reported by Moudgil {\it et al.}~\cite{ref:moudgil} (see references given therein to earlier STLS calculations).

In the present work we develop and evaluate an alternative approach to pair correlations in bilayer systems, in the absence of inter-layer tunnel and exchange, by means of an extension of the analytic theory given by Asgari {\it et al.}~\cite{ref:mehran_04} for the ground state of 2D quantum plasmas. This theory has provided a fully quantitative account of DMC data for the pair distribution function in monolayers of both electrons~\cite{ref:gorigiorgi_04} and charged bosons~\cite{ref:conti}, and has reproduced the location of the transition of the 2D electron gas from the paramagnetic to the spin-polarized state as reported in DMC work of Attaccalite {\it et al.}~\cite{ref:attaccalite}. In brief, the general strategy is to set up Schr\"{o}dinger-like differential equations for the ``pair amplitudes'' $\sqrt{g_{\alpha\beta}(r)}$ starting from a Fermi-hypernetted-chain (FHNC) approximation as first developed by Chakraborty and coworkers~\cite{ref:chakraborty}, which is then supplemented with in-plane three-body correlation terms~\cite{ref:wb_3} and tailored to embody the Hartree-Fock (HF) limit as well as a set of sum rules for a two-component fluid of charge carriers. The necessary input to account for inter-layer three-body terms is regretfully not available for a full test of the theory at the present time. The theory is free from the instability that appears in the quantum STLS approach of Moudgil {\it et al.}~\cite{ref:moudgil} in the strong-coupling regime.

The outline of the paper is briefly as follows. In Sect.~\ref{sect:theory} we present a formally exact zero-energy scattering equation for the pair amplitudes and introduce the approximations that we propose for the evaluation of the main contributions to the scattering potentials. In Sect.~\ref{sect:sum} we pause to show that these approximations satisfy the plasmon sum rule and the charge neutrality condition. Our numerical results for the pair distribution functions of EEB and EHB systems are presented in Sect.~\ref{sect:num} in comparison with the DMC simulation data~\cite{ref:rapisarda,ref:depalo_02,ref:moudgil}. Finally, Sect.~\ref{sect:concl} summarizes our main conclusions.

\section{THEORY}\label{sect:theory}

We consider a double-quantum-well system with electrons in the first well and electrons or holes in the second well, with masses $m_1$ and $m_2$ and areal densities $n_1$ and $n_2$. We assume the confining potential of each layer to be so high and narrow that we can neglect any interlayer tunneling and treat the motion of the carriers in the two wells as strictly two-dimensional. The distance $d$ between the layers is small so as to permit inter-layer interactions. The Hamiltonian of the system is
\begin{equation}
{\cal H}=-\frac{\hbar^2}{2}\sum_{(i,\alpha)}\frac{\nabla^2_{i,\alpha}}{m_{\alpha}} + \frac{1}{2}\sum_{\scriptscriptstyle (i,\alpha)\neq(j,\beta)}v_{\alpha\beta}(r_{ij})
\end{equation}
Here $v_{\alpha\alpha}=e^2/r$, $v_{\alpha\bar{\alpha}}=\pm e^2/\sqrt{r^2+d^2}$ with positive sign for EEB and negative sign for EHB, and $(i,\alpha)$ indicates the $i$-th particle in layer $\alpha$. Our aim is to calculate the pair distribution functions $g_{\alpha\beta}(r)$ of the bilayer, which are defined as
\begin{equation}
g_{\alpha\beta}(r)=\frac{1}{S}\frac{1}{n_{\alpha}n_{\beta}}\sum_{\scriptscriptstyle (i,\alpha)\neq(j,\beta)}\left\langle\delta({\bf r}-{\bf r}_{(i,\alpha)}+{\bf r}_{(j,\beta)})\right\rangle
\end{equation}
where the average is taken on the ground state and $S$ is the area of the sample.

We start the calculation, as in the work of Chakraborty and coworkers~\cite{ref:chakraborty} on binary boson mixtures and liquid metallic hydrogen, by introducing the two-component Jastrow-Slater variational theory involving the FHNC approximation. We make an {\it Ansatz} of the Jastrow-Feenberg form for the ground-state wave function as $\Psi=F_{11}F_{22}F_{12}\Phi_1\Phi_2$, where
\begin{equation}
\left\{
\begin{array}{l}
F_{\alpha\alpha}=\prod_{i<j}\exp \left\{\displaystyle \frac{1}{2}\left[u_{\alpha\alpha}(|{\bf r}_{(i,\alpha)}-{\bf r}_{(j,\alpha)}|)\right]\right\}
\vspace{0.2 cm}\\
F_{\alpha\bar{\alpha}}=\prod_{i,j}\exp \left\{\displaystyle \frac{1}{2}\left[u_{\alpha\bar{\alpha}}(|{\bf r}_{(i,\alpha)}-{\bf r}_{(j,\bar{\alpha})}|)\right]\right\}
\end{array}
\right.
\end{equation}
and $\Phi_{\alpha}$ is a Slater determinant of single-particle plane-wave states. The HNC closure relations~\cite{ref:feenberg} express $u_{\alpha\beta}$ in terms of $g_{\alpha\beta}$ as
\begin{equation}
u_{\alpha\beta}(r)=\ln g_{\alpha\beta}(r)-[g_{\alpha\beta}(r)-1]+C_{\alpha\beta}(r)\,,
\end{equation}
with $C_{\alpha\beta}(r)$ obeying in Fourier transform the relation
\begin{equation}
C_{\alpha\beta}(k)=S_{\alpha\beta}(k)-\delta_{\alpha\beta}-\sum_{\gamma}[S_{\alpha\gamma}(k)-\delta_{\alpha\gamma}]C_{\beta\gamma}(k).
\end{equation}
Here $S_{\alpha\beta}(k)=\delta_{\alpha\beta}+\sqrt{n_{\alpha}n_{\beta}}\int[g_{\alpha\beta}(r)-1]\exp{(i{\bf k}\cdot{\bf r})}d^2{\bf r}$ is the structure factor. Minimization of the ground-state energy leads to an Euler-Lagrange equation for the pair distribution functions,
\begin{equation}\label{EL}
\left[-\frac{\hbar^2}{2\mu_{\alpha\beta}}\nabla^2+V^{\rm eff}_{\alpha\beta}(r)\right]\sqrt{g_{\alpha\beta}(r)}=0
\end{equation}
where the Laplacian term derives from the von Weizs\"{a}cker-Herring ``surface'' kinetic energy~\cite{ref:herring}, $\mu_{\alpha\beta}$ is the reduced mass, and the effective scattering potential is given by
\begin{equation}\label{eff}
V^{\rm eff}_{\alpha\beta}(r)=v_{\alpha\beta}(r)+W_{\alpha\beta}^{\rm B}(r)+W_{\alpha\beta}^{\rm F}(r)\,.
\end{equation}
In Eq.~(\ref{eff}) the scattering potential is decomposed into the sum of three terms, where the Bose-like term $W_{\alpha\beta}^{\rm B}(r)$ contains the effect of correlations and by itself would determine the pair distribution functions in a boson fluid, whereas the Fermi term $W_{\alpha\beta}^{\rm F}(r)$ derives from exchange and from kinetic energy terms other than the surface kinetic energy.
	
The term $W^{\rm F}_{\alpha\beta}$ has a very complicated expression within the FHNC~\cite{ref:lantto}. However, in dealing with a one-component electron fluid Kallio and Piilo~\cite{ref:kallio} have proposed a simple and effective way to account for the antisymmetry of the fermion wave function. Their argument is immediately extended to a two-component Fermi fluid~\cite{ref:fhnc_3d} and leads to the requirement that the HF pair distribution functions are the solution of Eq.~(\ref{EL}) in the high-density regime in both layers. This prescription yields in Fourier transforms
\begin{equation}\label{fermi}
W^{\rm F}_{\alpha\alpha}(k)=\frac{\hbar^2}{m_{\alpha}}\int \frac{\nabla^2 \sqrt{g^{\rm HF}_{\alpha\alpha}(r)}}{\sqrt{g^{\rm HF}_{\alpha\alpha}(r)}}e^{i{\bf k}\cdot{\bf r}}d^2{\bf r}
-\lim_{r_s\to 0} W^{\rm B}_{\alpha\alpha}(k)
\end{equation}
and $W^{\rm F}_{\alpha{\bar\alpha}}(k)=0$. The rationale behind Eq.~(\ref{fermi}), which involves the intra-layer pair distribution function and structure factor in the HF approximation, is as follows. The first term on the RHS ensures that the HF limit is correctly embodied into the theory, and the second term ensures that the Bose-like scattering potential is suppressed for parallel-spin carriers at weak coupling.

Turning to the Bose-like scattering potential in Eq.~(\ref{eff}), its HNC expression has been derived by Chakraborty~\cite{ref:chakraborty} and its Fourier transform is given in terms of the structure factors of the system as
\begin{equation}\label{correlation}
\left\{
\begin{array}{l}
W^{\rm B}_{\alpha\alpha}(k)=-{\displaystyle\frac{\varepsilon_{\alpha}(k)}{2 n_{\alpha}}}\left\{2S_{\alpha\alpha}(k)-3+\left[S^2_{{\bar\alpha}{\bar\alpha}}(k)+\displaystyle\frac{m_{\alpha}}{m_{\bar \alpha}}S^2_{\alpha{\bar \alpha}}(k)\right]/\Delta^2(k)\right\}\vspace{0.2 cm}\\
W^{\rm B}_{\alpha{\bar \alpha}}(k)=-{\displaystyle \frac{\varepsilon_{\alpha}(k)S_{\alpha{\bar \alpha}}(k)}{2\sqrt{n_{\alpha}n_{\bar \alpha}}}}
\left\{1+\displaystyle\frac{m_{\alpha}}{m_{\bar \alpha}}-\left[S_{{\bar \alpha}{\bar \alpha}}(k)+\displaystyle\frac{m_{\alpha}}{m_{\bar \alpha}}S_{\alpha\alpha}(k)\right]/\Delta^2(k)\right\}
\end{array}
\right.
\end{equation}
where $\varepsilon_{\alpha}(k)=\hbar^2k^2/(2m_{\alpha})$ and $\Delta(k)=S_{11}(k)S_{22}(k)-S^2_{12}(k)$. Following Asgari {\it et al.}~\cite{ref:mehran_04} we have supplemented the HNC intra-layer potential in the first line of Eq.~(\ref{correlation}) by the inclusion of higher-order (three-body) Jastrow-Feenberg correlations, using the theory developed by Apaja {\it et al.}~\cite{ref:wb_3} with a weighting factor obtained by fitting DMC data on the ground state energy of the 2D charged-boson fluid~\cite{ref:conti}. Data for a similar inclusion of three-body terms in the inter-layer scattering potential are not available at present.

It is evident that the insertion of Eqs.~(\ref{eff})-(\ref{correlation}) into Eq.~(\ref{EL}) allows a self-consistent numerical calculation of the pair distribution functions and of the effective interactions. Before proceeding to the numerical solution of this problem, we examine how the approximate theory presented above fares in regard to some exact properties of pair correlations in bilayers.

\section{LIMITING BEHAVIORS AND PLASMON SUN RULE}\label{sect:sum}

We discuss in this Section how the present theory satisfies some exact properties that are characteristic of charged-particle fluids. The crucial point is that the HNC expression for the Bose-like scattering potential exactly cancels the bare Coulomb potential at large distance, all other terms in the scattering potential (including the three-body correlation term) being instead short-ranged. This property is proved through a careful analysis of the effective Schr\"{o}dinger equation in Eq.~(\ref{EL}) by following the line of argument presented in detail in the work of Davoudi {\it et al.}~\cite{ref:fhnc_3d}.

Considering for simplicity the case of a symmetric bilayer (with $n_1=n_2=n$ and $m_1=m_2=m$), inversion of Eqs.~(\ref{fermi}) and (\ref{correlation}) then yields the limiting behavior of the structure factors at long wavelengths as
\begin{equation}\label{lim}
S^{\rm eh}_{\alpha\beta}(k\rightarrow0)=(-1)^{|\alpha-\beta|}S^{\rm ee}_{\alpha\beta}(k\rightarrow0)
\rightarrow  \sqrt{\frac{\pi\hbar^2}{4m(Q_{11}-Q_{12})}}\frac{k}{k_{\rm F}}+ (-1)^{|\alpha-\beta|}\frac{\varepsilon_k}{2\hbar\omega_{\rm pl}}\,,
\end{equation}
where $k_{\rm F}=\sqrt{2\pi n}$ is the Fermi wave number, $\varepsilon_k=\hbar^2 k^2/(2m)$, $\omega_{\rm pl}=\sqrt{4 \pi n e^2k/m}$, and the $Q$'s are constants to be self-consistently evaluated for each type of bilayer. The form of Eq.~(\ref{lim}) immediately ensures that the charge neutrality condition
\begin{equation} 
n~\int d^2 {\bf r}~[g_{\alpha\beta}(r)-1]=-\delta_{\alpha\beta}
\end{equation}
is satisfied through the linear vanishing of the structure factors at long wavelengths.

Equation~(\ref{lim}) also ensures that the symmetric and antisymmetric plasmon sum rules are satisfied. These read
\begin{equation}\label{psr}
S_{\pm}(k\rightarrow0)\rightarrow \frac{\varepsilon_{k}}{\hbar \omega_{\pm}(k)}
\end{equation}
with $S_{\pm}\equiv S_{11}\pm S_{12}$ and
\begin{equation}\label{omega1}
\omega_{+}^{\rm eh}(k)=\omega_{-}^{\rm ee}(k)=\sqrt{\frac{k^2_{\rm F}(Q_{11}-Q_{12})}{4\pi m}}k\,,
\end{equation}
\begin{equation}\label{omega2}
\omega_{-}^{\rm eh}(k)=\omega_{+}^{\rm ee}(k)=\sqrt{\frac{4\pi ne^2k}{m}}\,.
\end{equation}
In the symmetric (antisymmetric) long-wavelength mode of the EHB (the EEB) the carriers in the two layers move in phase (out of phase) and screen each other's potential, leading to acoustic dispersion relations with coefficients determined by the asymptotic behavior of the screened potential~\cite{ref:dassarma_madhukar}.

\section{NUMERICAL RESULTS}\label{sect:num}

We turn to a presentation of our numerical results, which are obtained by solving Eq.~(\ref{EL}) through the following self-consistency cycle. We start with the trial choice $g_{\alpha\beta}(r)=g^{\rm HF}_{\alpha\beta}(r)$ and $W_{\alpha\beta}^{\rm B}(r)=0$, and evaluate the corresponding scattering potentials and structure factors. A Fourier transform of the latter leads to a new value of the input for $g_{\alpha\beta}(r)$ and $W_{\alpha\beta}^{\rm B}(r)$, and this procedure is continued till self-consistency between input and output is achieved.

\begin{figure*}
\begin{center}
\tabcolsep=0 cm
\begin{tabular}{cc}
\includegraphics[width=0.5\linewidth]{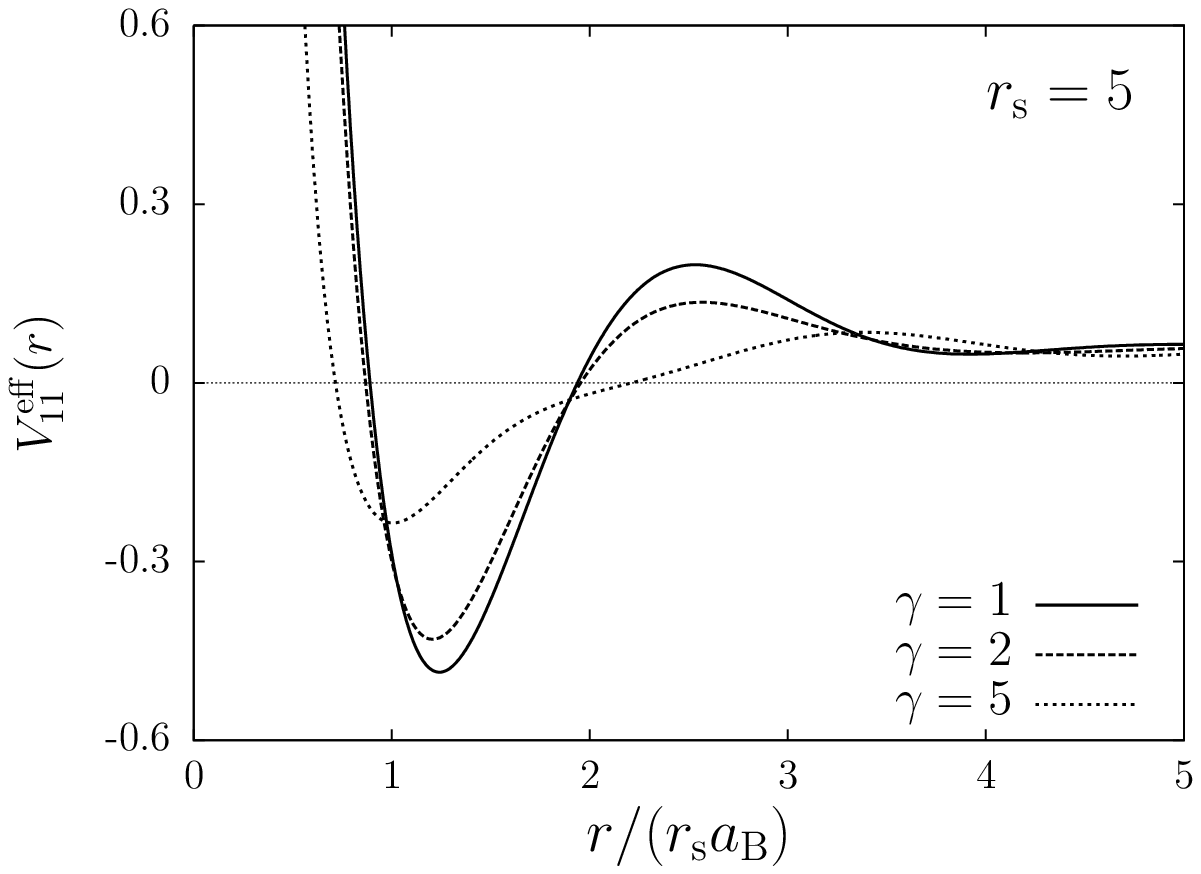}&
\includegraphics[width=0.5\linewidth]{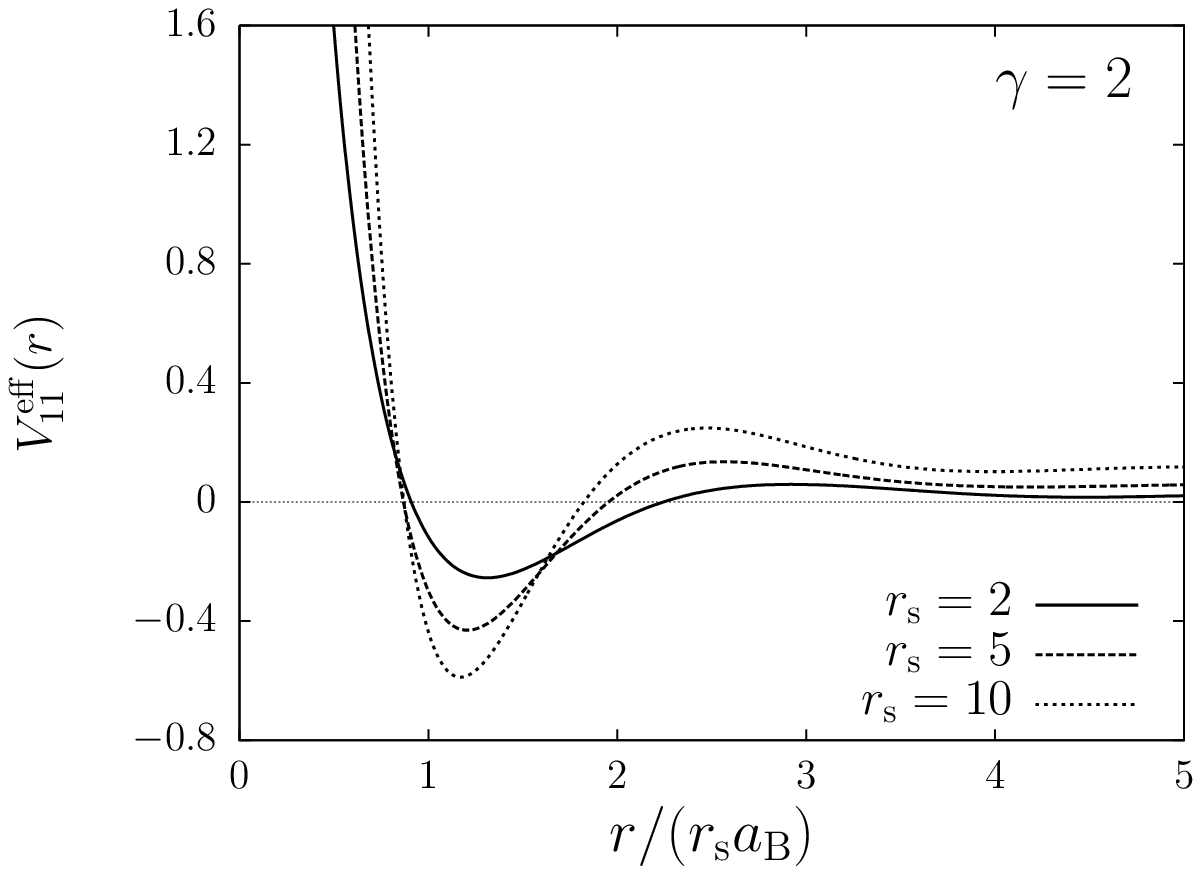}\\
\includegraphics[width=0.5\linewidth]{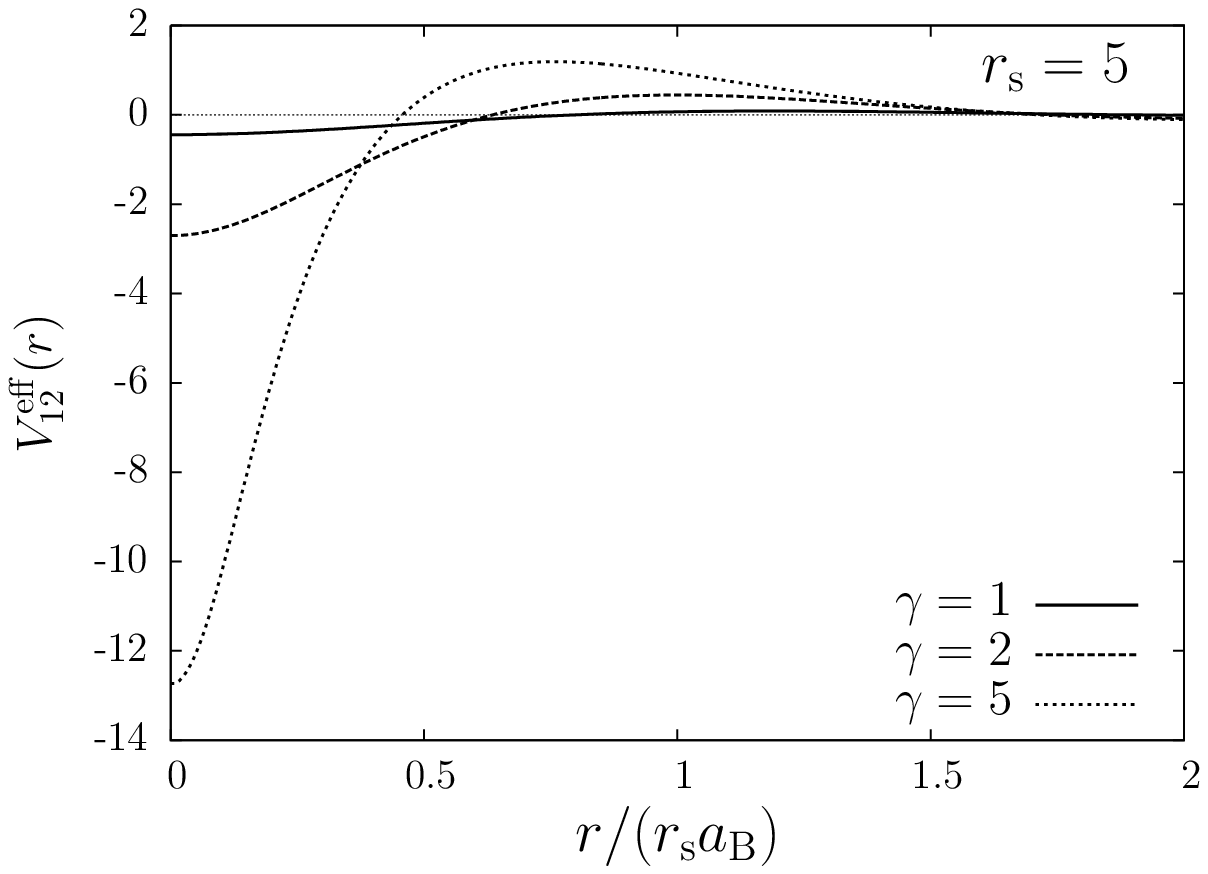}&
\includegraphics[width=0.5\linewidth]{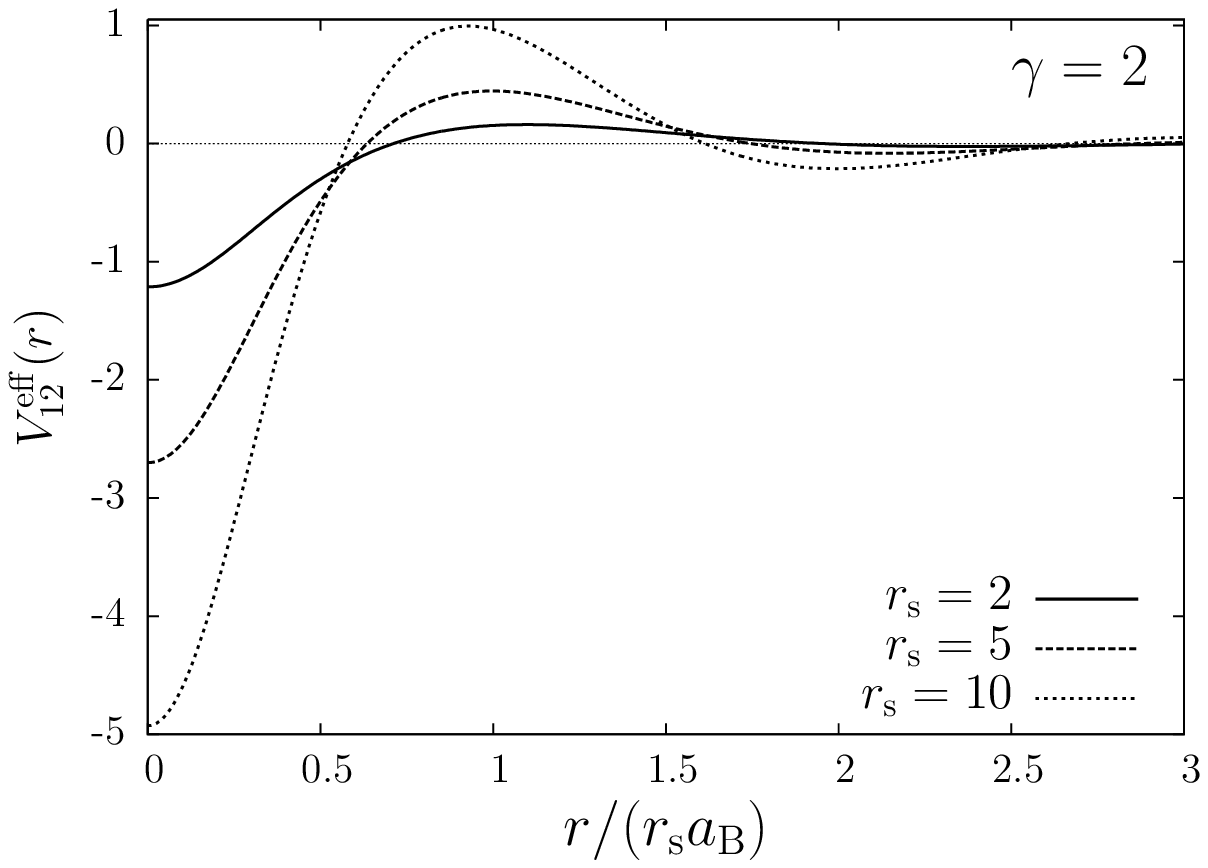}
\end{tabular}
\caption{Intra-layer (top) and inter-layer (bottom) scattering potentials $V^{\rm eff}_{\alpha\beta}(r)$ (in Rydberg units) as functions of $r/(r_s a_{\rm B})$ in symmetric electron-hole bilayers, for $r_s=5$ and varying inverse layer distance $\gamma=(r_s a_{\rm B})/d$ (left) and for varying $r_s$ at $\gamma=2$ (right).\label{fig1}}\end{center}
\end{figure*}

\begin{figure*}
\begin{center}
\tabcolsep=0 cm
\begin{tabular}{cc}
\includegraphics[width=0.5\linewidth]{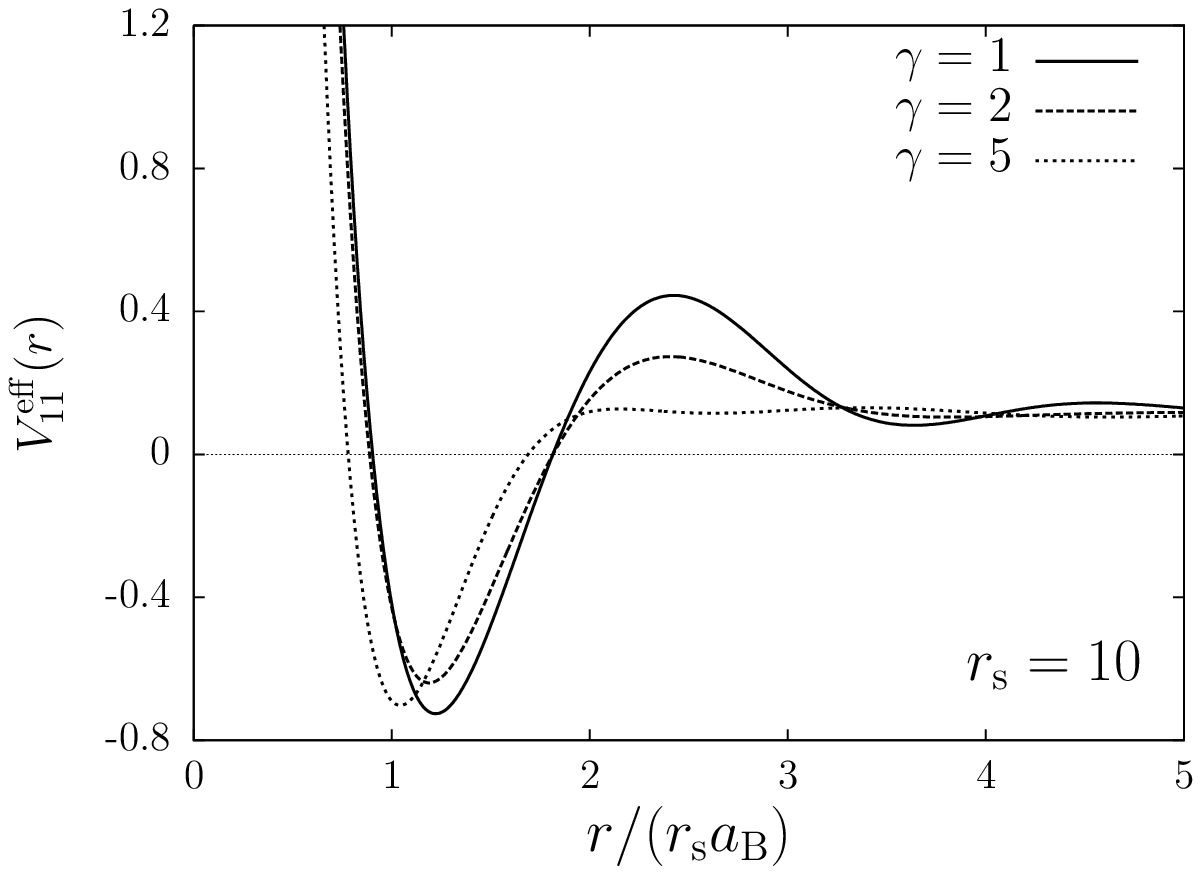}&
\includegraphics[width=0.5\linewidth]{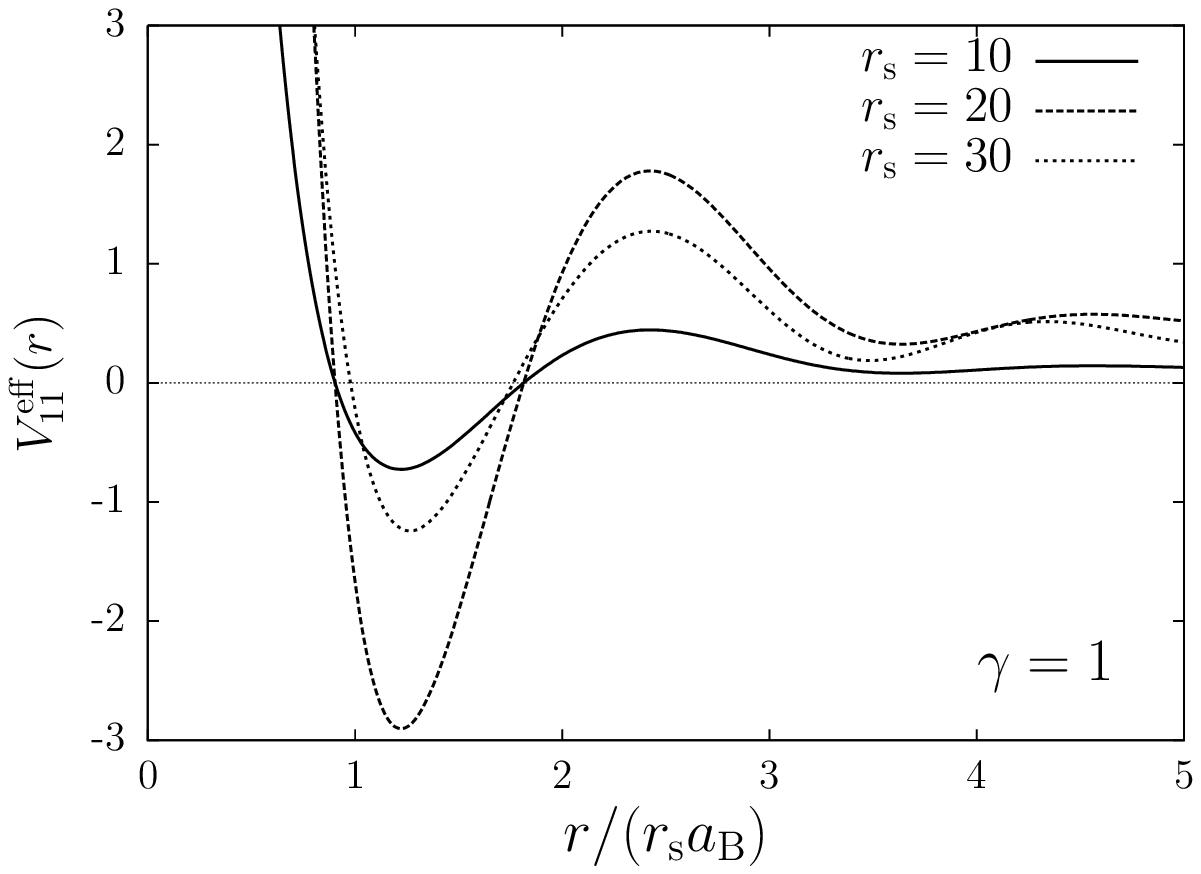}\\
\includegraphics[width=0.5\linewidth]{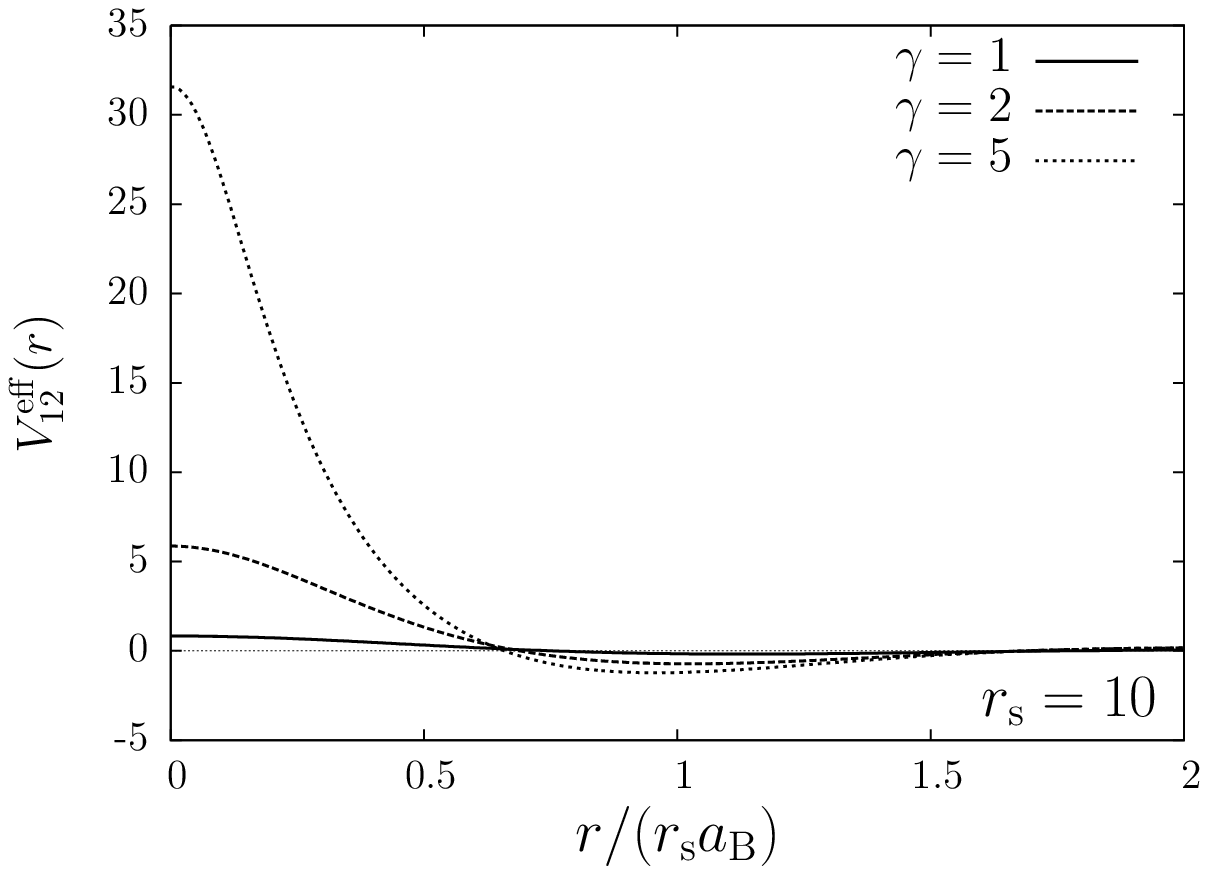}&
\includegraphics[width=0.5\linewidth]{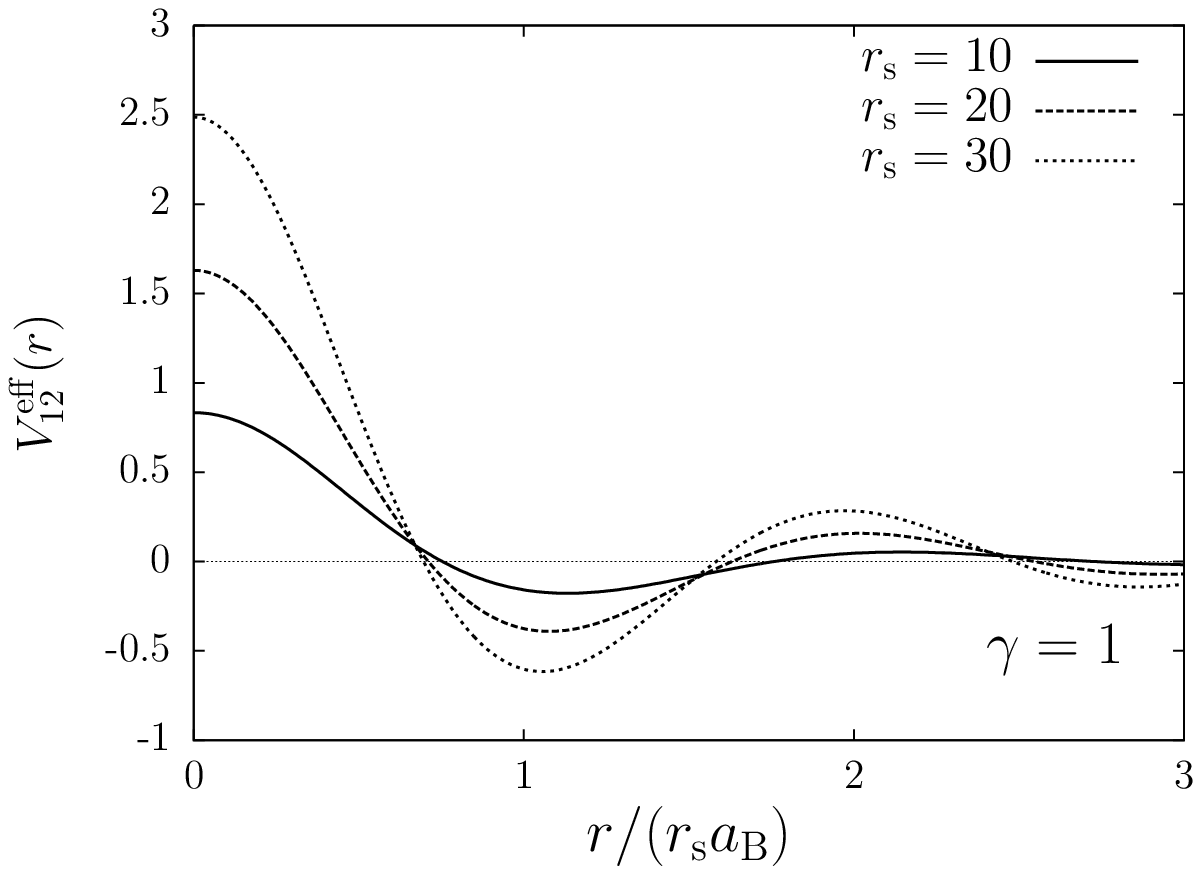}
\end{tabular}
\caption{Intra-layer (top) and inter-layer (bottom) scattering potentials $V^{\rm eff}_{\alpha\beta}(r)$ (in Rydberg units) as functions of $r/(r_s a_{\rm B})$ in symmetric electron-electron bilayers, for $r_s=10$ and varying inverse layer distance $\gamma=(r_s a_{\rm B})/d$ (left) and for varying $r_s$ at 
$\gamma=1$ (right).\label{fig2}}\end{center}
\end{figure*}

\begin{figure*}
\begin{center}
\tabcolsep=0 cm
\begin{tabular}{cc}
\includegraphics[width=0.5\linewidth]{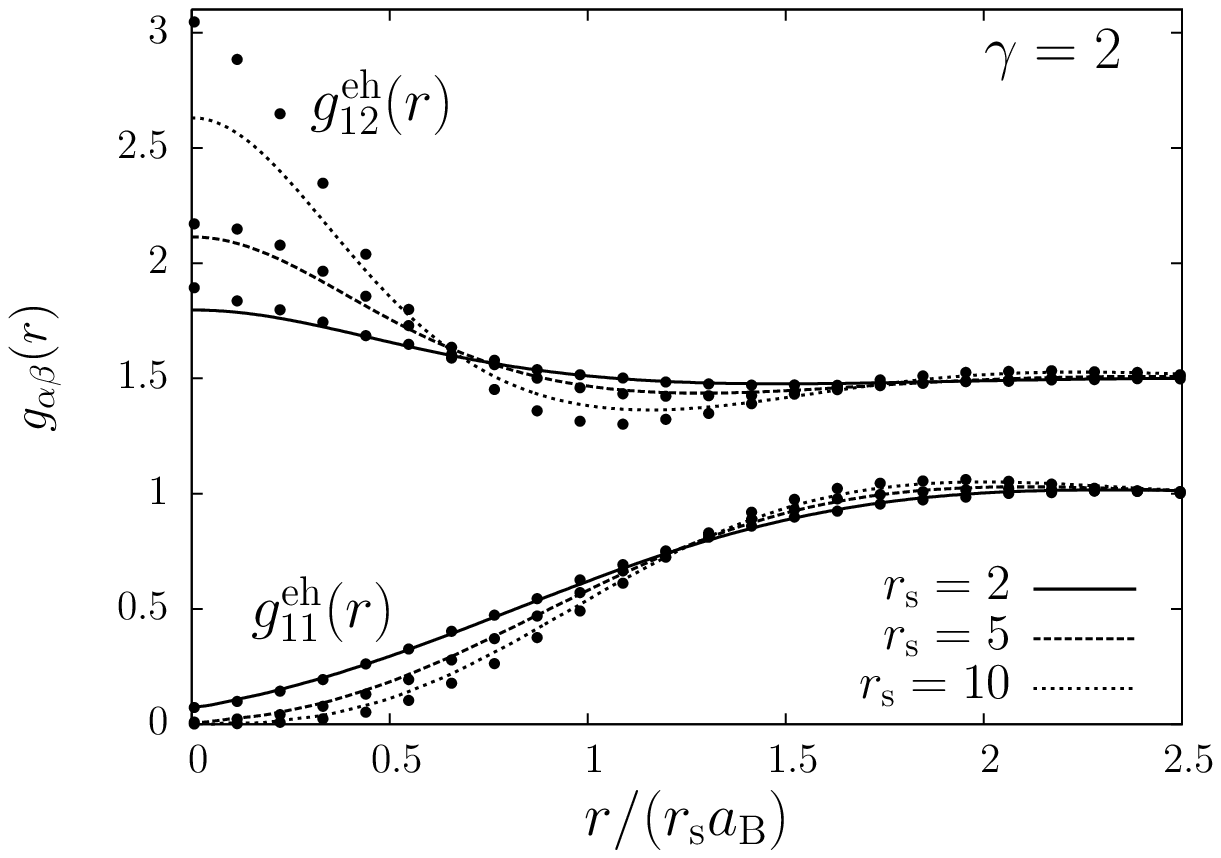}&
\includegraphics[width=0.5\linewidth]{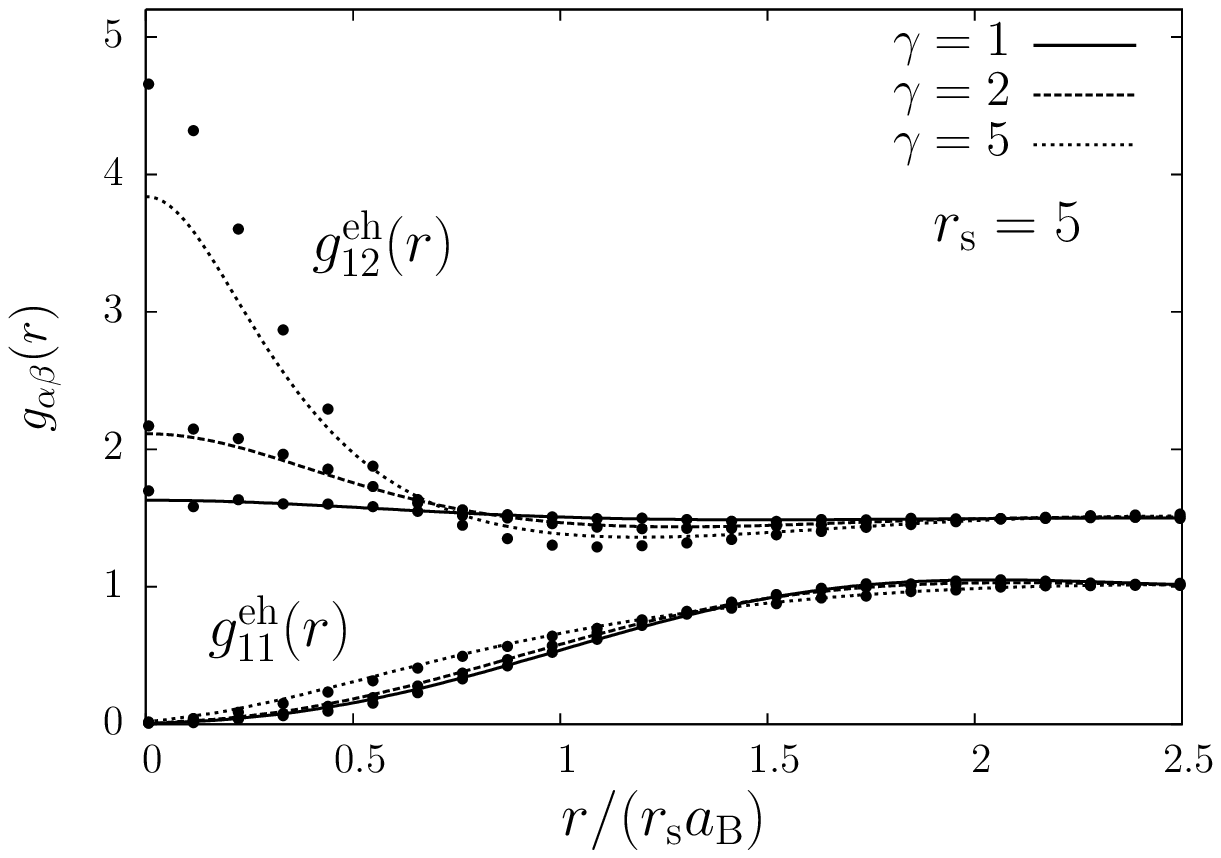}\\ 
\includegraphics[width=0.5\linewidth]{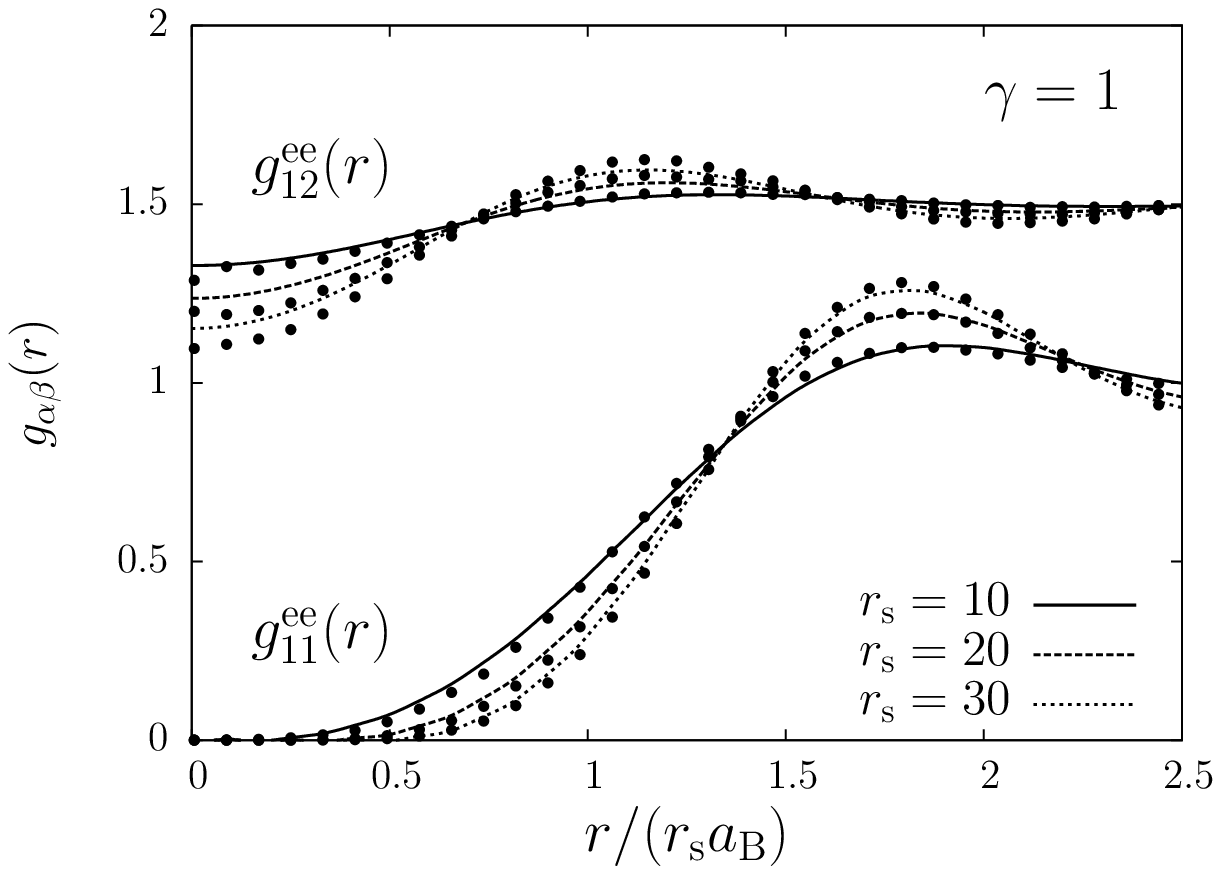}&
\includegraphics[width=0.5\linewidth]{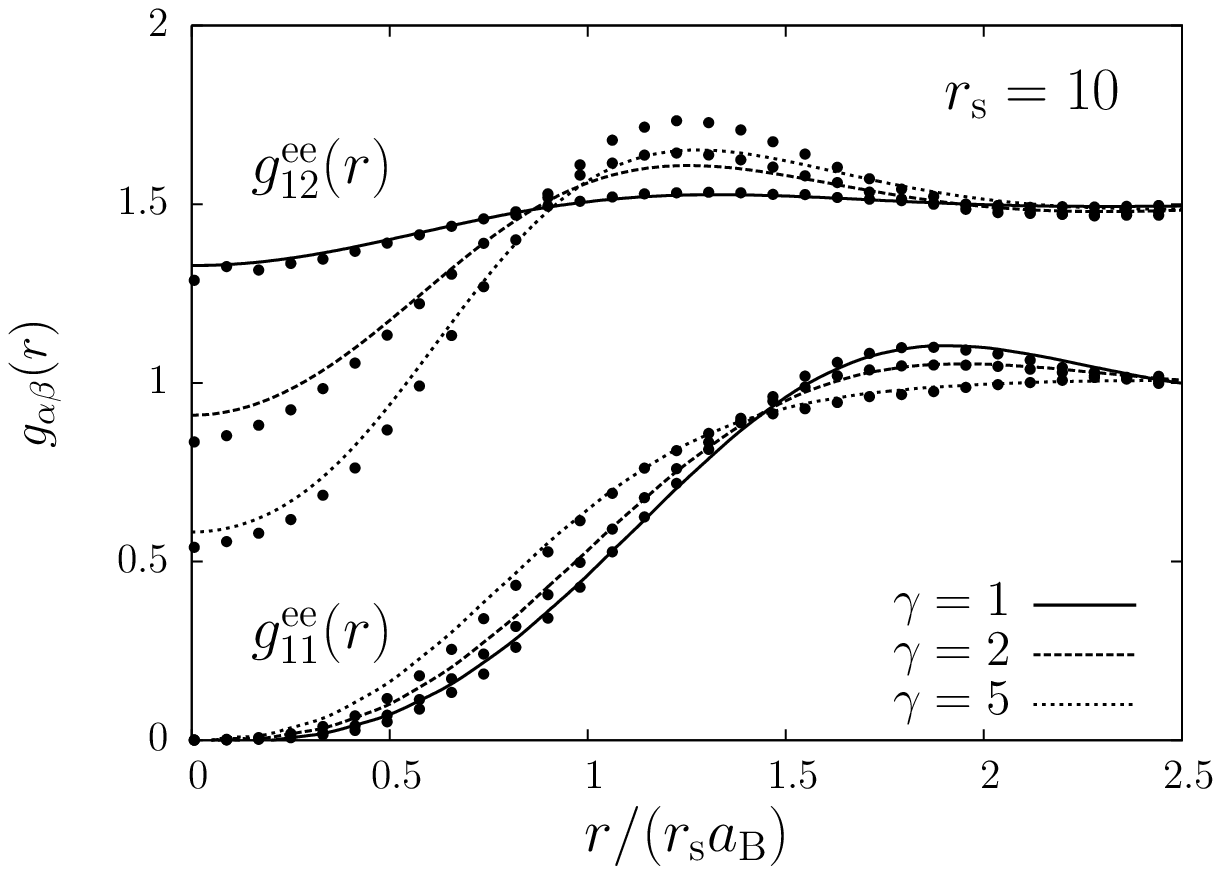}
\end{tabular}
\caption{Pair distribution functions $g_{\alpha\beta}(r)$ as functions of $r/(r_s a_{\rm B})$ in symmetric electron-hole bilayers (top) and electron-electron bilayers (bottom), for varying $r_s$ at fixed inverse layer distance $\gamma=(r_s a_{\rm B})/d$ (left) and varying $\gamma$ at fixed $r_s$ (right). The results for $g_{12}(r)$ have been shifted upwards by $0.5$ for clarity. The dots report data from DMC simulation runs~\cite{ref:rapisarda,ref:moudgil,ref:depalo_02}.\label{fig3}}
\end{center}
\end{figure*}

Figures~\ref{fig1} and \ref{fig2} illustrate the dependence of the effective scattering potentials on the Coulomb coupling strength $r_s$ and on the inverse inter-layer spacing $\gamma=(r_s a_{\rm B})/d$ for EHB and EEB systems, respectively. The relative phases of the oscillations shown by these functions in the two layers immediately indicate that inter-layer short-range order is growing with either of these two parameters, this state of order being in-phase for EHB systems and out-of-phase for EEB systems. This result, which is in full accord with the available DMC data~\cite{ref:rapisarda,ref:depalo_02,ref:moudgil}, is illustrated in Figure~\ref{fig3}.

The degree of agreement between theory and DMC data is illustrated in Figure~\ref{fig3}. The inclusion of three-body correlations in the calculation of the effective intra-layer potentials yields almost fully quantitative agreement with the data, as was the case for the monolayer system studied by Asgari {\it et al.}~\cite{ref:mehran_04}. In the absence of such three-body terms, the FHNC theory shows some  quantitative discrepancies from the data in the inter-layer correlations, and such discrepancies are becoming severe in the EHB case as the two layers are brought closer together. The limit in which inter-layer excitons are formed in the electron-hole bilayer is evidently incorrectly described by the simple FHNC theory.

\section{CONCLUSIONS}\label{sect:concl}

In summary, we have presented a self-consistent analytical theory of intra-layer and inter-layer two-body correlations in electron-hole and electron-electron bilayers, and compared its results with data from Quantum Diffusion Monte Carlo runs. The FHNC theoretical approach yields quantitatively valuable results for intra-layer short-range order when it is supplemented with three-body correlation terms in the construction of the scattering potentials. There appears to be a growing need for similar three-body effects in the calculation of inter-layer correlations with growing inter-layer coupling. Incorporation of the limit of inter-layer exciton formation should be sought for electron-hole bilayers.

\ack 
We are indebted to S. de Palo and G. Senatore for providing us with their Quantum Monte Carlo data reported in our figures, and to F. Capurro and B. Davoudi for early contributions to the numerical work.

\end{document}